\documentclass{PoS}
\usepackage{xspace}
\usepackage{amsmath}
\title{Measurement of tW the production cross-section at 13 TeV with CMS}

\ShortTitle{}

\author{\speaker{Sergio S\'anchez Cruz}\thanks{Partially funded by Becas Severo Ochoa del Principado de Asturias.} on behalf of the CMS Collaboration\\
        Universidad de Oviedo\\
        E-mail: \email{sergio.sanchez.cruz@cern.ch}}

\abstract{The inclusive cross-section for tW production in proton-proton collisions at $\sqrt{s} = 13$ TeV is measured with a dataset corresponding to an integrated luminosity of 35.9 fb$^{-1}$ collected by the CMS experiment. The measurement is performed in events with one electron and one muon, and exploits kinematic differences between the signal and the dominating $t\bar{t}$ background through the use of multivariate discriminants designed to separate the two processes. The measured cross-section of $\sigma = 63.1 \pm 1.8~({\rm stat}) \pm 6.0~({\rm syst}) \pm 2.1~({\rm lumi})$ pb is in agreement with standard model expectations. }

\FullConference{Sixth Annual Conference on Large Hadron Collider Physics (LHCP2018)\\
                4-9 June 2018\\
                Bologna, Italy}

\newcommand{\ttbar}{\ensuremath{{\mathrm{t}\overline{\mathrm{t}}}}\xspace}
\newcommand{\muObs}{\ensuremath{0.88 \pm 0.02\text{ (stat.) } \pm 0.09\text{ (syst.) } \pm 0.03\text{ (lumi.) }}\xspace}       
\newcommand{\resultxsecmain}{\ensuremath{63.1 \pm 1.8\text{ (stat.) } \pm 6.4\text{ (syst.) } \pm 2.1\text{ (lumi.) }} pb\xspace}  

\begin{document}

\section{Introduction}

Single top quarks are produced via the electroweak interaction. There are three main production 
channels that can occur in proton-proton (pp) collisions: the exchange of a virtual W boson ($t$ channel), 
the production and decay of a virtual W boson ($s$ channel) and the associated production of a 
top quark and a W boson (tW channel). 

The tW process was first observed by the CMS~\cite{Chatrchyan:2014tua} and ATLAS~\cite{Aad:2015eto} Collaborations in 8 TeV
pp collisions. This process provides a unique opportunity to study the standard model and its 
extensions, due to the quantum interference of this process with top quark pair production (\ttbar). 
It can also be a probe of the $V_{tb}$ element of the CKM matrix. 

The increase in cross-section of the process at 13 TeV and the larger amount of luminosity delivered 
by the LHC to the CMS detector~\cite{CMS} allows the measurement of the tW production-cross section with a 
larger precision than ever. This note reports that measurement, fully described in~\cite{CMS-TOP-17-018}.

\section{Event selection and signal extraction} 

In the standard model (SM), a top quark decays almost exclusively into a W boson and a bottom quark. 
Therefore, the measurement is performed in events with two opposite-sign, different-flavor leptons,
targeting the leptonic decays of the two W bosons. Such events 
are collected using a set of di-lepton and single-lepton triggers. The leading lepton
is required to have $p_T >$ 25 GeV, while the subleading is required to have $p_T >$ 20 GeV.

\begin{figure}
\centering
\includegraphics[width=0.6\textwidth]{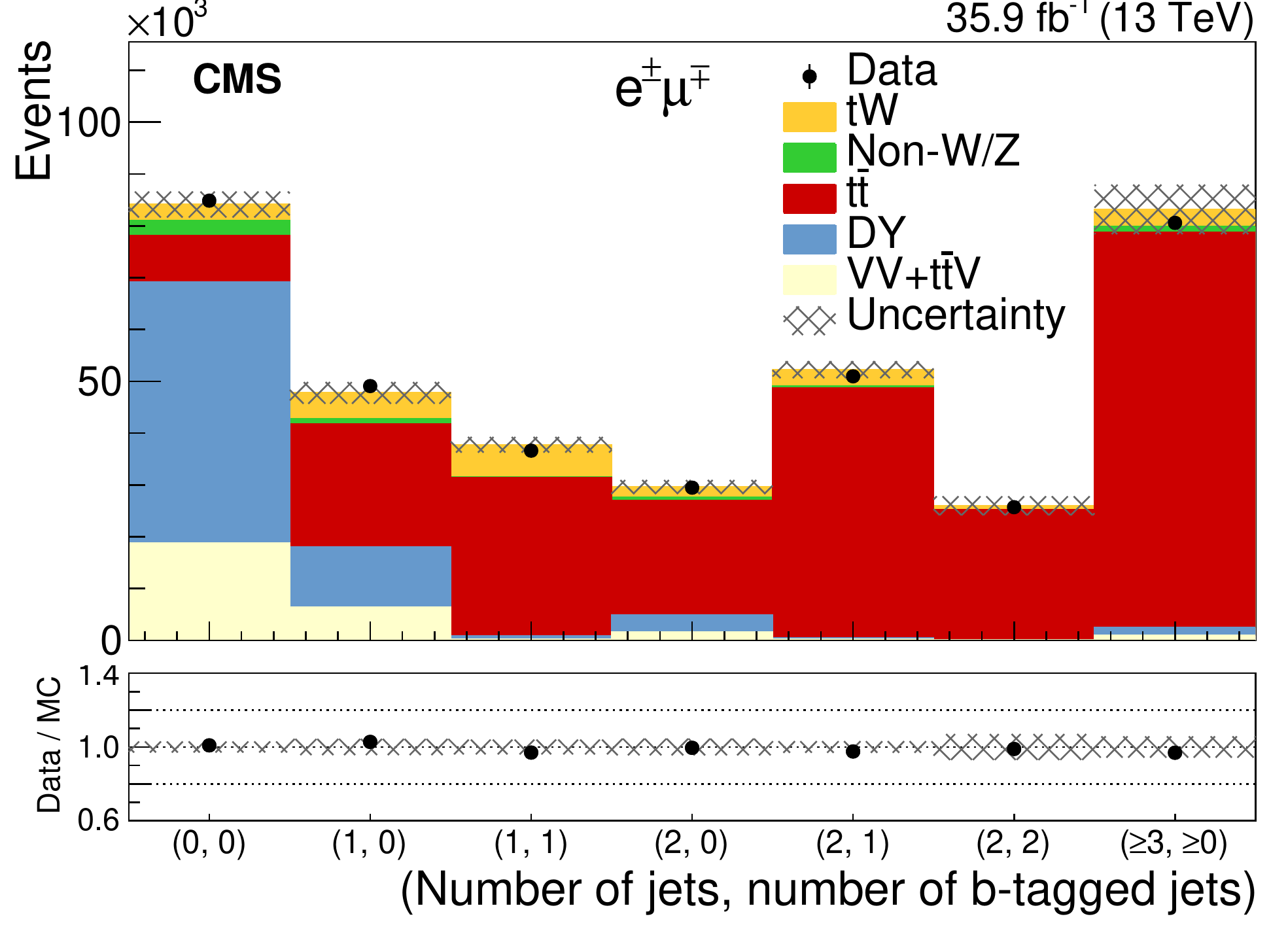}
\caption{Distribution of $(n_{\text{jet}}, n_{\text{b jet}})$ in events passing the di-leptonic 
selection~\cite{CMS-TOP-17-018}.}
\label{fig:njetnb}
\end{figure}

The analysis exploits the different $(n_{\text{jet}}, n_{\text{b jet}})$ spectrum
between the signal and the different backgrounds present in the dileptonic baseline selection. 
The distribution is shown in Fig.~\ref{fig:njetnb}, that shows that indeed the Drell-Yan 
events contain less jets or b jets, while \ttbar and tW events contain at least one b-tagged
jet. Three analysis regions are defined according to they different signal and background
composition: a signal-enriched region with events with exactly one jet that is b-tagged (1j1b), 
and two background-dominated regions with two jets, one with one b-tagged jet (2j1b) and one 
with two b-tagged jets (2j2b). 

Even the more signal-enriched region is still dominated by \ttbar events. Therefore, given 
there is no single observable that discriminates between tW and \ttbar events, the analysis
makes use of multivariate techniques in order to achieve some discrimination.

\begin{figure}[htbp!]                                                                                                
  \centering   
    \includegraphics[width=0.49\textwidth]{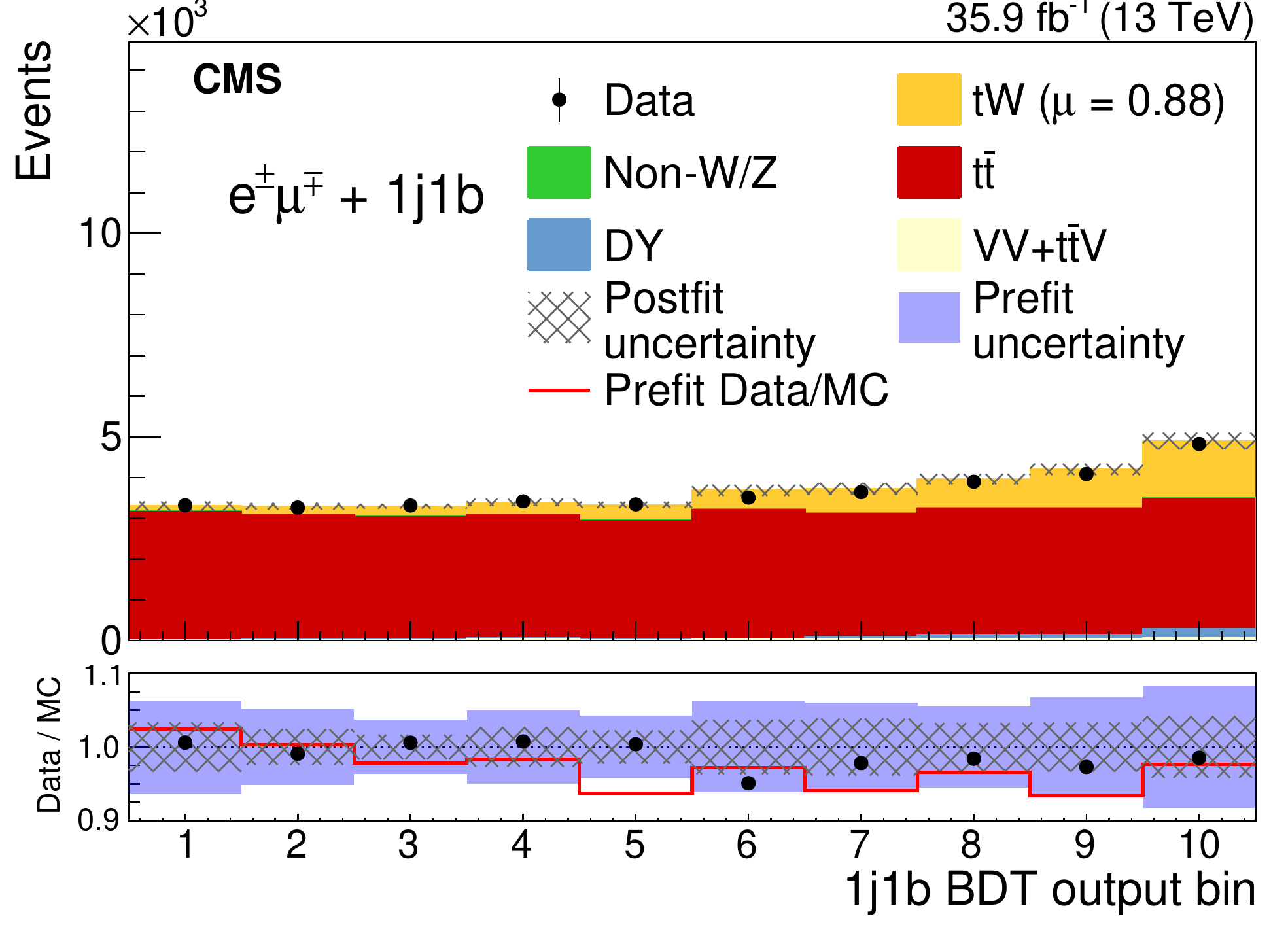}   
    \includegraphics[width=0.49\textwidth]{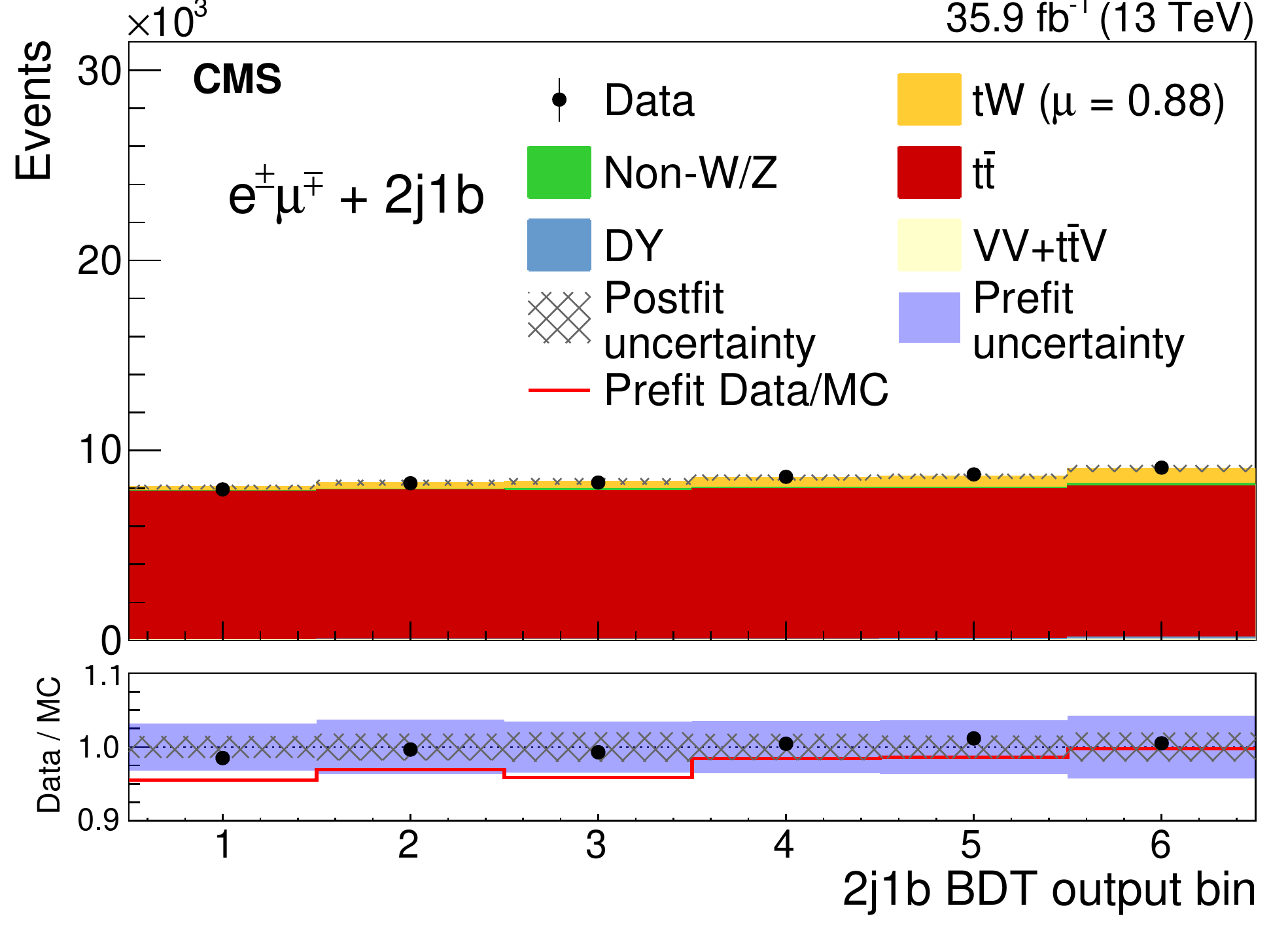} \\
    \includegraphics[width=0.49\textwidth]{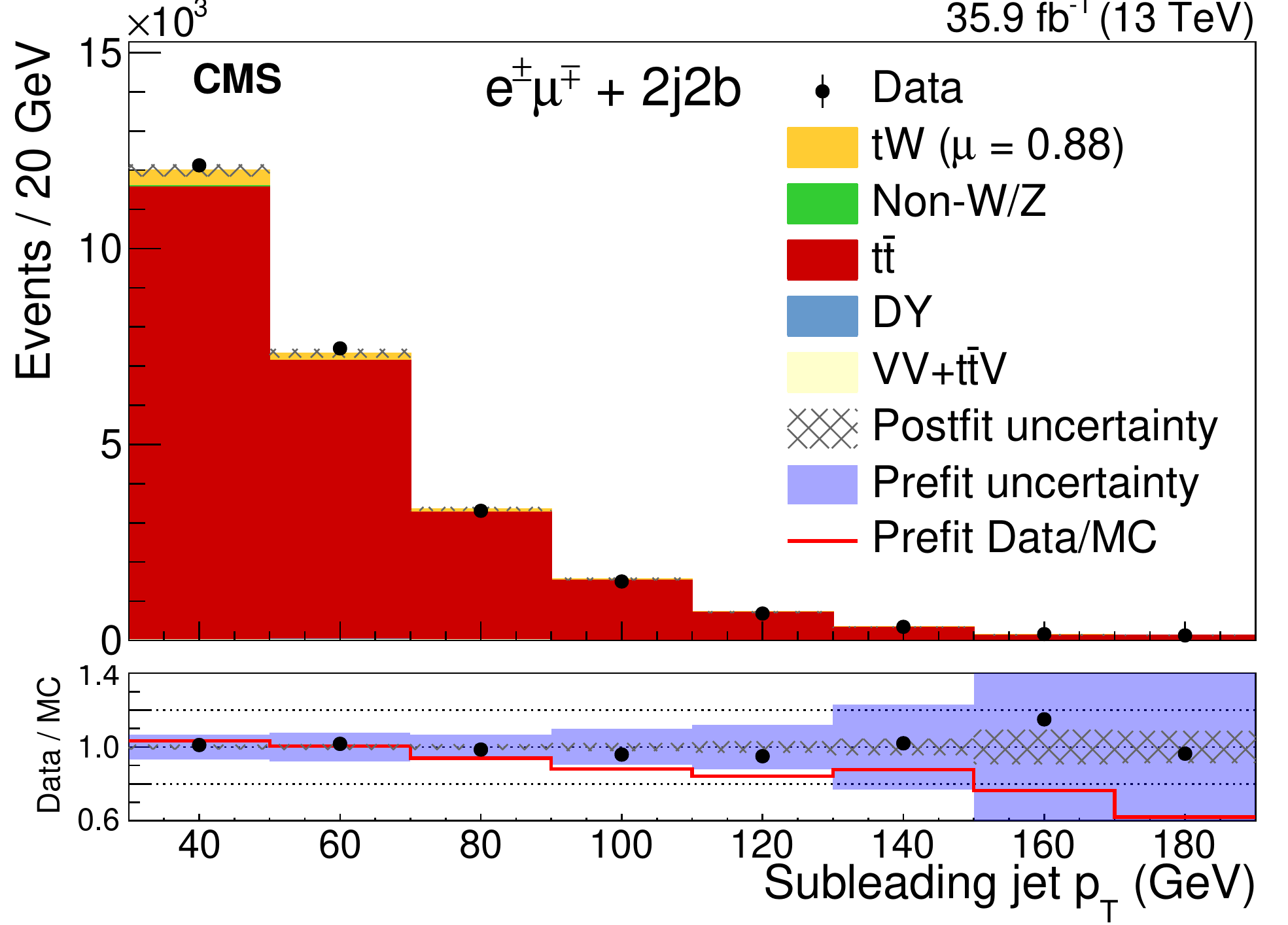}   
\caption{Comparison of the BDT output in the 1j1b (upper left) and 2j1b (upper right) regions and the $p_{T}$ of the subleading jet    
in the 2j2b region (lower) distributions after the fit is performed for the observed data and simulated events~\cite{CMS-TOP-17-018}. }
\label{fig:fig_outputBDTdataPostFit}   
\end{figure}

Dedicated boosted decision trees (BDTs) are trained in the 1j1b and 2j1b regions, exploiting 
the different kinematic and topological differences between \ttbar and tW production. The 
BDT trained in the 1j1t region profits from the fact that in \ttbar events in this region
one of the jets lies outside the acceptance. Therefore important variables used in the 
training of this BDT are the number of jets with $p_T > 20$ GeV or the overall transverse 
boost of the system formed by the two leptons, the jet and the missing transverse momentum.
The BDT used in the 2j2b region exploits angular correlations between the leptons, 
the jet and the missing transverse momentum. 

The signal extraction is done by performing a maximum likelihood fit to the distribution
of the two BDTs in the 1j1b and 2j2b regions. Additionally, the distribution of the 
subleading jet $p_T$ in events in the 2j2b region is also included in the fit, which
is performed simultaneously in the three regions. The distributions employed in the fit 
are shown in Fig.~\ref{fig:fig_outputBDTdataPostFit}. 

The statistical analysis employs a likelihood function, $\mathcal{L}(\mu, \vec{\theta})$, 
which is a function of the signal strength, $\mu$, and a set of nuisance parameters 
that model the effect of the systematic uncertainties present in the analysis. The likelihood
function is built as the product of Poisson distributions that represent the event
counting in each one of the bins of the fitted distributions. Additional log-normal
terms are added modeling the priors of each nuisance parameter. 

The best estimation of $\mu$ is obtained by minimizing  $\mathcal{L}(\mu, \vec{\theta})$
with respect to all of its parameters. The 68\% confidence interval is obtained 
by considering variations of the test statistic used in~\cite{fitcosas} by one unit
from its minimum. 

The impact of the statistical uncertainty is determined from the uncertainty of a fit
performed fixing all the nuisance parameters to the postfit value of the nominal fit. 
The impact of the systematic uncertainty is assessed by performing a likelihood fit
fixing all the nuisance parameters to the postfit value of the nominal fit, except 
the one under study.

The systematic uncertainties taken into account are listed in Tab.~\ref{tab:systs_fit}
and account for both experimental uncertainties, as well as theoretical assumptions 
that are made on the distributions of the signal process and the normalization
and distribution of the background processes.

\section{Results} 

The estimated tW signal-strength parameter is \muObs, corresponding to a measured 
cross-section of \resultxsecmain, consistent with the SM expectations with an uncertainty of 11\%. The impact of each source of systematic uncertainty in the 
fit is shown in Table~\ref{tab:systs_fit}. The uncertainties in the luminosity, trigger and lepton efficiencies have a sizable impact in the  estimation of the background yield, which is dominant in all bins of the fit. Therefore these uncertainties have a significant contribution in the final measurement.

\begin{table}[htbp]  
\centering \caption{Estimation of the effect on the signal strength of each source of uncertainty in the fit. Exp\
erimental and modeling uncertainties affect both the rate and the shape of the templates while background normalizat\
ion uncertainties affect only the rate~\cite{CMS-TOP-17-018}.}     
\label{tab:systs_fit}
\vspace{0.2cm}
\begin{tabular}{lc}  \hline\hline
    Source & Uncertainty (\%)  \\ \hline   \hline
Experimental &   \\     \hline 
 Trigger efficiencies  & 2.7 \\   
   Electron efficiencies  & 3.2 \\    
 Muon efficiencies  & 3.1 \\    
   JES  & 3.2 \\    
   Jet energy resolution  & 1.8 \\    
     b tagging efficiency  & 1.4 \\    
 Mistag rate  & 0.2 \\    
Pileup  & 3.3 \\ \hline \hline
Modeling & \\  \hline
  \ttbar\ $\mu_\mathrm{R}$ and $\mu_\mathrm{F}$ scales & 2.5 \\     
  tW $\mu_\mathrm{R}$ and $\mu_\mathrm{F}$ scales & 0.9 \\   
 Underlying event  & 0.4 \\     
   Matrix element/PS matching  & 1.8 \\     
Initial-state radiation  & 0.8 \\     
Color reconnection & 2.0 \\     
     B fragmentation  & 1.9 \\     
Semileptonic B decay  & 1.5 \\     
 PDFs  & 1.5 \\     
DR-DS  & 1.3 \\ \hline \hline
Background normalization & \\\hline  
    \ttbar   & 2.8 \\     
  VV   & 0.4 \\     
Drell--Yan   & 1.1 \\     
   Non-W/Z leptons & 1.6 \\     
  $\ttbar$V  & 0.1 \\
MC finite sample size & 1.6 \\
   Full phase space extrapolation & 2.9 \\
     Total systematic & 10.1  \\    
Integrated luminosity & 3.3 \\
Statistical & 2.8 \\
Total & 11.1 \\    
\hline \hline 
\end{tabular} 
\end{table}


\begin{thebibliography}{99}
\bibitem{Chatrchyan:2014tua}
CMS Collaboration, ``Observation of the associated production of a single top quark and a $W$ boson in pp collisions at $\sqrt s ~=~ $8 TeV.'' \textit{ Phys. Rev. Lett.} \textbf{112} (2014) 231802

\bibitem{Aad:2015eto}
ATLAS Collaboration, ``Measurement of the production cross-section of a single top quark in association with a $W$ boson at 8 TeV with the ATLAS experiment'', \textit{JHEP} \textbf{01} (2016) 064

\bibitem{CMS}
CMS Collaboration, JINST 3 S08004 (2008)

\bibitem{CMS-TOP-17-018}
CMS Collaboration, ``Measurement of the production cross section for single top quarks in association with W bosons in proton-proton collisions at $\sqrt{s} = 13$ TeV'',  JHEP \textbf{10} (2018) 117

\bibitem{fitcosas} 
CMS Collaboration, ``Precise determination of the mass of the Higgs boson and tests of compatibility of its couplings with the standard model predictions using proton collisions at 7 and 8 $\,\text {TeV}$'', Eur. Phys. J. C 75 (2015) 212

\end{thebibliography}
\end{document}